# Automating Iconclass:
# LLMs and RAG for Large-Scale Classification of Religious Woodcuts

Drew B. Thomas
University College Dublin

*Abstract:*

This paper presents a novel methodology for classifying early modern religious images by using Large Language Models (LLMs) and vector databases in combination with Retrieval-Augmented Generation (RAG). The approach leverages the full-page context of book illustrations from the Holy Roman Empire, allowing the LLM to generate detailed descriptions that incorporate both visual and textual elements. These descriptions are then matched to relevant Iconclass codes through a hybrid vector search. This method achieves 87% and 92% precision at five and four levels of classification, significantly outperforming traditional image and keyword-based searches. By employing full-page descriptions and RAG, the system enhances classification accuracy, offering a powerful tool for large-scale analysis of early modern visual archives. This interdisciplinary approach demonstrates the growing potential of LLMs and RAG in advancing research within art history and digital humanities.

## 1. Introduction

The advent of the printing press in Early Modern Europe revolutionized the dissemination of graphic design and visual media.[1] For the first time, illustrations were reproduced at an unprecedented scale, appearing in millions of copies of books across the continent. These illustrations played a crucial role in shaping public opinion, educating readers, and conveying religious and political messages. However, despite their importance, studying the content of these images at scale has remained a daunting task. The absence, until recently, of a comprehensive, systematic record of which early modern books contain illustrations has further complicated efforts to analyze them in meaningful ways.

The "Visualizing Faith: Print, Piety, and Propaganda" project at University College Dublin addresses this gap by focusing on one of the most significant uses of visual media during this period: the Protestant Reformation. This was Europe's first mass media event, where both Protestants and Catholics strategically employed printed images to educate their followers,

1. This publication has emanated from research conducted with the financial support of Taighde Éireann – Research Ireland, under Grant number 21/PATH-A/9655 at University College Dublin. This work was first presented at the "Digital Humanities and Artificial Intelligence" conference at the University of Reading on 17 June 2024.

reinforce theological principles, and denigrate their opponents. Religious illustrations were thus essential tools for conveying ideology and shaping public perception during a time of intense religious conflict and transformation.

While previous scholarship has often focused on analyzing a small number of high-profile images or specific texts, this paper explores a method to study these illustrations at scale by developing a technique for automatically applying Iconclass codes—a controlled vocabulary system widely used to classify visual content in cultural heritage—to religious and biblical illustrations from the Holy Roman Empire.[2] Iconclass, developed in the 1940s in the Netherlands by Henri van de Waal, offers a structured, hierarchical system that transitions easily from narrow to broader queries, making it a valuable tool for cataloguing and interpreting historical images.[3] However, manually assigning Iconclass codes to large datasets is labor-intensive and time-consuming, which underscores the importance of automating this process. The primary objective of this experiment was to develop and evaluate a method for automatic classification of early modern religious images, focusing on two of Iconclass's ten base categories: "Religion and Magic" and "Bible."

> 7. Bible
>
> 73. New Testament
>
> 73D. Passion of Christ
>
> 73D2 the episode of the Last Supper
>
> 73D23 Christ washes the feet of the apostles (John 13:1-20)
>
> 73D231 Christ washes Peter's feet

**Fig. 1: Iconclass Example**

Recent advances in multi-modal Large Language Models (LLMs), combined with improvements in semantic embeddings using vector databases, present new opportunities for examining the visual content of historical images. This paper introduces a novel methodology that leverages these tools. By feeding an image into an LLM, a detailed textual description of the image is produced. This description can then be queried against a vector database of pre-existing Iconclass descriptions, allowing for a highly accurate classification of the image's content. The method is tested using several different model configurations, and the results show a significant improvement over traditional keyword-based or image classification methods. This approach achieves an average precision of 87% at five levels of Iconclass classification, marking a substantial leap in the ability to analyze visual media from the early modern period.

---

2. https://iconclass.org/
3. Henri van de Waal, *Iconclass an iconographic classification system*, completed and edited by L.D. Couprie with R.H. Fuchs, E. Tholen & G. Vellekoop (North-Holland, Amsterdam, 1974-1985).

In this paper, I evaluate the performance of these models and demonstrate how they can be applied to large image corpora to enable new forms of research in history, art history and the digital humanities. This method not only provides an efficient solution for image classification but also highlights the growing role of LLMs and vector databases in automating tasks traditionally dominated by manual classification. Through this interdisciplinary approach, we move closer to unlocking the vast visual archives of early modern Europe, offering new insights into the role of religious and political imagery during one of the most transformative periods in European history.

## 2. Related Work

Recent advancements in object detection and image captioning have demonstrated significant potential in automating the classification and analysis of cultural heritage data. Vision-language models fine-tuned on custom datasets have shown that deep learning can generate meaningful descriptions in art-historical contexts.[4] Several projects have applied computer vision techniques for the study of early modern books. The Oxford Broadside Ballad project used an object detector to identify woodcuts in early English ballads.[5] Similarly, the Compositor project used

---

4. Eva Cetinic, "Iconographic Image Captioning for Artworks," *arXiv*, February 7, 2021. https://doi.org/10.48550/arXiv.2102.03942.
5. Giles Bergel, et al. "Content-Based Image-Recognition on Printed Broadside Ballads: The Bodleian Libraries'

computer vision to build a large database of printers' ornaments.[6] The Ornamento project used a similar methodology to identify illustrations and ornamentation in books printed across Europe and the Spanish Americas up to the year 1600.[7] Such projects were foundational to building collections of images for further investigation into the content of printed artwork, a task also addressed with computer vision.

However, several studies have noted the complexity of applying these techniques to art historical data, which often lacks the large-scale, richly annotated datasets necessary for training sophisticated models. Cultural heritage datasets pose challenges, particularly in terms of object detection and the difficulties of generalizing models trained on modern, natural images to artworks that depict symbolic or non-realistic subjects.[8] Despite these limitations, approaches combining computer vision and metadata analysis, such as in the DEArt dataset, have shown promise in improving the precision of object detection in heritage domains by incorporating external knowledge sources like Wikidata.[9] Other projects, such as OmniArt, iART, SMKExplore and MINERVA, have created large image collections to use as benchmarks or platforms to present such collections.[10]

Object detection seeks to increase the searchability and discoverability of image

---

ImageMatch Tool," IFLA WLIC 2013 - Singapore - Future Libraries: Infinite Possibilities, 2017. https://library.ifla.org/id/eprint/209

6. Hazel Wilkinson, et al. "Computer Vision and the Creation of a Database of Printers' Ornaments," *Digital Humanities Quarterly,* 15, no. 1 (May 21, 2021).
7. Alexander S. Wilkinson, "Ornamento Europe: Towards an Atlas of the Visual Geography of the Renaissance Book," in Arthur der Weduwen and Malcolm Walsby (eds.), *The Book World of Early Modern Europe* (Leiden: Brill, 2022), pp. 547–62.
8. Artem Reshetnikov, et al., "DEArt: Dataset of European Art," *arXiv*, November 3, 2022. https://doi.org/10.48550/arXiv.2211.01226.
9. Ibid.
10. Gjorgji Strezoski and Marcel Worring, "OmniArt: Multi-Task Deep Learning for Artistic Data Analysis," *arXiv*, August 2, 2017. https://doi.org/10.48550/arXiv.1708.00684 ; Gjorgji Strezoski and Marcel Worring, "OmniArt: A Large-Scale Artistic Benchmark," *ACM Trans. Multimedia Comput. Commun. Appl.*, 14, no. 4 (October 23, 2018), pp. 1-21. https://doi.org/10.1145/3273022 ; Matthias Springstein, et al., "iART: A Search Engine for Art-Historical Images to Support Research in the Humanities," in *Proceedings of the 29th ACM International Conference on Multimedia* (New York: Association for Computing Machinery, 2021), pp. 2801–3. https://doi.org/10.1145/3474085.3478564 ; Louie Meyer, et al., "Algorithmic Ways of Seeing: Using Object Detection to Facilitate Art Exploration," in *Proceedings of the 2024 CHI Conference on Human Factors in Computing Systems* (New York: Association for Computing Machinery, 2024), pp. 1–18. https://doi.org/10.1145/3613904.3642157 ; Matthia Sabatelli, et al., "Advances in Digital Music Iconography: Benchmarking the Detection of Musical Instruments in Unrestricted, Non-Photorealistic Images from the Artistic Domain," *Digital Humanities Quarterly*, 15, no. 1 (March 5, 2021).

collections. One project developed a model to identify the captions of illustrations in early modern books which can then be made searchable.[11] Others use it to examine the images themselves, such as identifying specific objects in religious art to identify specific saints[12] or to classify illustrations in nineteenth-century children's literature.[13] However, using models trained on modern images can introduce errors. By applying temporal metadata to the classes detected by a model, one project learned to filter out anachronistic objects (e.g. a TV will not appear in an early modern painting).[14] But other projects, such as SniffyArt, are interested in more abstract topics. They trained a model to identify smell gestures in historical artwork to use in a hybrid model for smell-gesture recognition.[15] Some projects even seek to identify social concepts, such as revolution, violence or friendship,[16] while others have applied it to identify problematic images in colonial photographic archives.[17] Recently, the "Saint George on a Bike Project" used object detection to increase accuracy in caption generation.[18]

In the context of Iconclass, several studies have aimed to automate its application to art collections. Using CLIP, a state-of-the-art vision-language model, the Iconclass project itself has recently introduced semantic and image search into their online interface. This provides a framework that blends manual classification with machine learning.[19] Building on this, recent

---

11. Julia Thomas and Irene Testini, "Capturing Captions: Using AI to Identify and Analyse Image Captions in a Large Dataset of Historical Book Illustrations," *Digital Humanities Quarterly,* 18, no. 2 (2024). https://digitalhumanities.org/dhq/vol/18/2/000740/000740.html.
12. Federico Milani and Piero Fraternali, "A Dataset and a Convolutional Model for Iconography Classification in Paintings," *J. Comput. Cult. Herit.,* 14, no. 4 (July 16, 2021), pp. 1-18. https://doi.org/10.1145/3458885.
13. Yongho Kim, et al., "Applying Computer Vision Systems to Historical Book Illustrations: Challenges and First Results," 2021. https://www.researchgate.net/publication/356843079
14. Maria-Cristina Marinescu, et al., "Improving Object Detection in Paintings Based on Time Contexts," in *2020 International Conference on Data Mining Workshops* (2020), pp. 926–32. https://doi.org/10.1109/ICDMW51313.2020.00133.
15. Mathias Zinnen, et al., "SniffyArt: The Dataset of Smelling Persons," in *Proceedings of the 5th Workshop on analySis, Understanding and proMotion of heritAge Contents*, SUMAC '23 (New York: Association for Computing Machinery, 2023), pp. 49–58. https://doi.org/10.1145/3607542.3617357.
16. Delfina Pandiani, et al., "Automatic Modeling of Social Concepts Evoked by Art Images as Multimodal Frames," *arXiv* (October 14, 2021). https://doi.org/10.48550/arXiv.2110.07420.
17. Jonathan Dentler, et al., "Sensitivity and Access: Unlocking the Colonial Visual Archive with Machine Learning," *Digital Humanities Quarterly,* 18, no. 2 (2024). https://digitalhumanities.org/dhq/vol/18/2/000742/000742.html.
18. Saint George on a Bike. https://saintgeorgeonabike.eu/
19. Cristian Santini, et al., "Multimodal Search on Iconclass Using Vision-Language Pre-Trained Models," in *Proceedings of the 2023 ACM/IEEE Joint Conference on Digital Libraries* (Santa Fe, New Mexico: IEEE Press, 2024), pp. 285–87. https://doi.org/10.1109/JCDL57899.2023.00061 ; Etienne Posthumus, Harald Sack, and Hans Brandhorst, "The Art Historian's Bicycle Becomes an E-Bike," in *6th Workshop on Computer VISion for ART* research has shown that certain visual concepts in Iconclass are easier to learn than others, highlighting differences that could guide future improvements in text-to-image similarity learning.[20] One project focusing on accurately predicting Iconclass codes presents a multi-modal retrieval system, combining textual and visual features to improve classification accuracy through transfer learning.[21] While these methods show great potential, they also underscore the reliance on fine-tuning neural networks to achieve optimal performance, a computationally intensive process.

    Fine-tuning has been a common thread in many of these applications. However, these fine-tuned, image-focused approaches face limitations in domains where textual descriptions play a critical role in classification. While textual features outperform visual features in some Iconclass retrieval tasks, the combination of both provides the best results.[22] There is growing recognition that hybrid models, which incorporate both text and image, may offer the best solutions to cultural heritage classification challenges. Yet even these approaches still depend on deep learning models and feature extraction techniques that are resource-intensive.

    In contrast to the above methodologies, this project diverges by using LLMs for generating textual descriptions of images, relying less on image-based models and more on the semantic power of text. Inspired by the multimodal approaches, which implement vision-language models like CLIP for Iconclass classification, this methodology uses OpenAI's GPT-4o model to generate detailed descriptions based on both visual and textual elements without the need for fine-tuning or computationally heavy CNNs. This method enables more flexibility in applying classifications, especially for complex cultural heritage objects that rely heavily on

---

*Analysis In Conjunction with the 2022 European Conference on Computer Vision* (Tel Aviv, Israel: 2022).
20. Kai Labusch and Clemens Neudecker, "Gauging the Limitations of Natural Language Supervised Text-Image Metrics Learning by Iconclass Visual Concepts," in *Proceedings of the 7th International Workshop on Historical Document Imaging and Processing* (New York: Association for Computing Machinery, 2023), pp. 19–24. https://doi.org/10.1145/3604951.3605516.
21. Nikolay Banar, Walter Daelemans, and Mike Kestemont, "Transfer Learning for the Visual Arts: The Multi-Modal Retrieval of Iconclass Codes," *J. Comput. Cult. Herit.,* 16, no. 2 (June 24, 2023). https://doi.org/10.1145/3575865.
22. Nikolay Banar, Walter Daelemans, and Mike Kestemont, "Multi-Modal Label Retrieval for the Visual Arts: The Case of Iconclass," in *Proceedings of the 13th International Conference on Agents and Artificial Intelligence* (Vienna, Austria:

interpretive context.

## 3. Corpus

The image corpus used in this study was built by the Ornamento project at University College Dublin. Ornamento is a comprehensive archive of visual elements from early modern books. The project, consisting of Prof. Alexander Wilkinson and myself, employed a CNN trained using the YOLOv3 algorithm, to detect and classify pages containing woodcut illustrations and other forms of ornamentation.[23] We then used the VGG Image Search Engine (VISE) to group together impressions of the same woodcuts used in different books.[24] To allow for the searching across visual content, we implemented the WISE Image Search Engine,[25] which embedded images with the vision-language model OpenCLIP and created a search index based on approximate nearest neighbor.[26] This allows us to search the visual content of images with a text search query—even in the absence of keyword tagging—or by uploading an image to find similar results. In total, this yielded a substantial dataset, comprising approximately six million visual items extracted from nearly seventy million pages from 200,000 digital scans of books published before 1601.[27]

The subset of Ornamento used for "Visualizing Faith" is limited to images from the Holy Roman Empire classified as illustrations within religious or biblical texts. This selection process has resulted in a corpus of approximately 120,000 images.

For the purposes of this article, the test set is composed of three distinct groups:

---

SCITEPRESS - Science and Technology Publications, 2021), pp. 622–29.
23. Joseph Redmon and Ali Farhadi, "Yolov3: An incremental improvement." *arXiv preprint arXiv:1804.02767* (2018).
24. Abhishek Dutta, Relja Arandjelović, and Andrew Zisserman, *VGG Image Search Engine* (2021). https://www.robots.ox.ac.uk/~vgg/software/vise/.
25. Prasanna Sridhar, et al., "WISE Image Search Engine (WISE)," *Wiki Workshop (10th Edition)* (2023). https://gitlab.com/vgg/wise/wise
26. Mehdi Cherti, et al., "Reproducible Scaling Laws for Contrastive Language-Image Learning," in *2023 IEEE/CVF Conference on Computer Vision and Pattern Recognition* (2023), pp. 2818–29. https://doi.org/10.1109/CVPR52729.2023.00276.

1. **Illustrations from Martin Luther's 1534 Bible**: This group includes 117 illustrations from the 1534 edition of Luther's Bible, his first German translation of the complete Bible and a foundational text of the Reformation. I used a published list of the illustrations with captions to apply Iconclasses for the ground-truth.[28] The images used come from the copy held at the Berlin State Library.[29]

2. **Illustrations from Martin Luther's 1551 Bible**: This group comprises 165 illustrations from the 1551 edition of Luther's Bible. The Löhe Memorial Library at Australian Luther University owns a copy acquired in 1954.[30] They issued a catalogue of the illustrations with captions which formed the basis of the ground truth.[31] The printer, Hans Lufft, produced four versions of the Bible in 1551. The images in this study come from a different edition from the Australian copy that used the same woodcuts. The images used in the project come from a copy from the Württembergische Landesbibliothek in Stuttgart.[32]

3. **Selected Thematic Illustrations**: The third group consists of images that represent key biblical narratives, classified according to the following iconographic themes:

    a. Adam and Eve (60 images)

    b. Noah's Ark (26 images)

    c. The Annunciation (49 images)

    d. The Nativity (45 images)

    e. The Last Supper (72 images)

---

27. Wilkinson, "Ornamento Europe."
28. Carl C. Christensen, "Luther and the Woodcuts to the 1534 Bible," *Lutheran Quarterly,* 19, no. 4 (Winter 2005), pp. 392–413.
29. *Biblia das ist die gantze heilige schrifft Deudsch* (Wittenberg: Hans Lufft, 1534). Berlin, Staatsbibliothek, 4" Bu 9401. VD16 B 2694. USTC 616653.
30. *Biblia das ist: die gantze heilige schrifft: Deudsch. auffs new zugericht* (Wittenberg: Hans Lufft, 1551 [1550]). Adelaide, Australian Luther College, RB CB72 1551. VD16 B 2730. USTC 616495.
31. *Luther Bible Images Catalogue: Illustrations from the 1551 Luther Bible* (Adelaide: Australian Lutheran College, 2022). See also, Trevor Schaefer, "Luther Bible, 1551 Edition," *Lutheran Theological Journal,* 49, no. 3 (November 9, 2020): pp. 171–78.
32. *Biblia: das ist die gantze heilige schrifft: Deudsch* (Wittenberg: Hans Lufft, 1551). Stuttgart, Württembergische Landesbibliohtek, Bb deutsch 155007. VD16 B 2729. USTC 616664. The illustrations in this edition are the same

    f. The Crucifixion (59 images)

These thematic images were identified using the vision-language model applied to Ornamento, as described above. The themes (e.g. "Adam and Eve") where used as search queries, which ensured an easy and quick way of identifying images for the test set. This approach allowed for precise identification and categorization of iconographically significant illustrations within the broader dataset. Thus, the entire test corpus consists of both detailed biblical illustrations that appear only once and more familiar themes that are depicted multiple times. This was done deliberately to test the model's consistency, ensuring it applies similar classifications to similar images across multiple instances of the same scene.

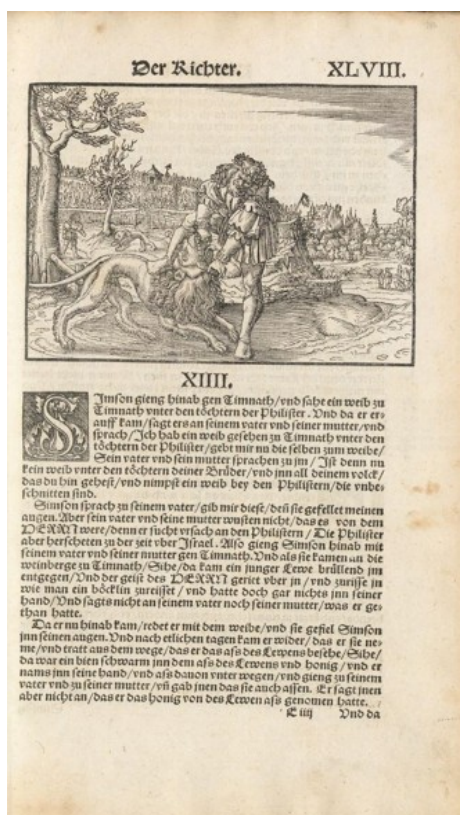

***Fig. 2:*** *A page from Luther's Complete Bible from 1534 depicting Samson wrestling a lion.*

**4. Methodology**

**4.1 Image-Based Search**

---

woodcuts as the Australian edition except for one or two exchanges. These were rectified prior to the analysis.

To establish a baseline for comparison, I queried my images with the Iconclass AI Test Set,[33] distributed by Iconclass and sampled from the Arkyves database.[34] The original test set contains 87,749 images, each associated with specific Iconclass codes, as listed in an accompanying JSON file. I filtered this dataset to include only Iconclass codes beginning with 1 or 7 ("Religion and Magic" and "Bible"), which reduced the set to 21,422 images.

I embedded the images using OpenCLIP and performed an exhaustive nearest-neighbor search for direct image-to-image comparisons.[35] The project images were used as queries in this vision-language model, which retrieved the most visually similar images from the filtered Iconclass AI Test Set. For retrieval, I implemented a method similar to the recently introduced Iconclass+ interface, which performs a nearest neighbor search, totals up the Iconclass codes of the nearest neighbors, and presents them in decreasing order.[36] For my implementation, I retrieved the top 10 nearest neighbors for each query image, counted the Iconclass codes assigned to each image, and used the most frequent code as the predicted classification.

### 4.2 Text-Based Search

### 4.2.1 Image Description Generation

While the image-search method relies on visual cues to identify relevant Iconclass codes, the text-based method leverages the descriptive power of LLMs. To generate descriptive text for each image in the corpus, I used the multi-modal capabilities of GPT-4o, which at the time of analysis was OpenAI's state-of-the-art large language model.[37] Each image was inserted via the API into a prompt that indicated the image was a woodcut illustration from an early modern Bible or religious book. Two distinct types of descriptions were generated for each illustration:

---

33. Iconclass AI Test Set, https://iconclass.org/testset/
34. Hans Brandhorst, "A Word Is Worth a Thousand Pictures: Why the Use of Iconclass Will Make Artificial Intelligence Smarter." https://iconclass.org/testset/ICONCLASS_and_AI.pdf. For the Arkyves database, see https://www.arkyves.org/
35. Thanks to the Irish Centre for High-End Computing for providing access to the Meluxina supercomputer for this task.
36. Santini, et al., "Multimodal Search on Iconclass Using Vision-Language Pre-Trained Models." See also,

- **Full-Page Descriptions**: These descriptions were generated using the full page from the book, allowing the LLM to consider surrounding text, headings, chapter numbers, and other contextual elements that might influence the interpretation of the image.
- **Illustration Descriptions**: In this case, the LLM generated descriptions based solely on the cropped illustration, devoid of any surrounding textual context. This approach tests the LLM's ability to interpret the visual content of the image in isolation.

**4.2.2 Database Construction**

The purpose of the image descriptions is to query them against a database of Iconclass descriptions. For this, I constructed a vector database using the open source database Weaviate, which stores both objects and vectors.[38] The database includes all hierarchical classifications under the "Religion and Magic" and "Bible" categories, totaling about 12,600 possible classifications.[39] This database supports various search approaches, including traditional keyword search, vector search, hybrid search, and Retrieval-Augmented Generation (RAG).

To accommodate different levels of classification detail, I created two versions of the database:

- **Basic Iconclass Database**: This version contains the standard Iconclass classifications, offering a straightforward list of codes and their descriptions. (e.g. "71B32: the building of the ark, and the embarkation (Genesis 7:5-9)")
- **Hierarchical Iconclass Database**: This version extends the basic database by including descriptions that reflect the full hierarchy of each classification, adding contextual information from parent levels in the system. (e.g. "71B32: Bible; Old Testament;

---

Posthumus, "The Art Historian's Bicycle Becomes an E-Bike."
37. The specific model was gpt-4o-2024-05-13.
38. https://github.com/weaviate/weaviate
39. Only the descriptions, and not the codes, were embedded in the vector database.

> Genesis from the descendants of Cain and Seth to Abraham; story of Noah; the building of the ark, and the embarkation (Genesis 7:5-9)")

Notice in the example that there is no mention of Noah in the basic database description but in the hierarchical database, Noah's name appears.

### 4.2.3 Keyword Search

To establish a baseline for text-based search, I employed a traditional keyword search. The descriptions were searched in both the basic and hierarchical databases, relying on direct textual matching between the image description and the Iconclass entries. This search process used the BM25 (Best Match 25) algorithm, a popular information retrieval method for ranking documents based on their relevance to a query. BM25 operates by calculating a score for each Iconclass entry based on the presence and frequency of keywords in the descriptions, as well as how rare those terms are across the database. The algorithm then returns the Iconclass entries with the highest BM25F scores—those deemed most relevant to the query description. In essence, this approach relies purely on keyword matching, providing a baseline for comparison with more advanced search methods.

### 4.3 Advanced Search Methods

After testing the keyword-based approach, I explored more advanced search methods to improve classification accuracy by incorporating semantic understanding and hybrid techniques. Vector embeddings are numerical arrays that represent objects, in this case, the Iconclass descriptions, in a way that captures their underlying meaning. Generated using OpenAI's text embedding model, these embeddings transform text into a format that allows machine learning models to analyze and compare them.[40] The meaning of each value in the array depends on the model used, and similarity between objects is judged by comparing their vector values using a distance metric. For this project, I used Weaviate's default metric, cosine distance, which measures the angle between two vectors to determine how closely they align.

### 4.3.1 Vector Search

Using embeddings for a vector search allows for a deeper comparison between the descriptions and the vectorized Iconclass classifications, capturing more abstract similarities beyond simple keyword matching. For example, in a traditional keyword search, a query for "Holy Communion" will not include results containing "Eucharist", whereas a vector search would recognize the semantic similarity and retrieve both sets of results. The vector search identified the most semantically similar Iconclass entries and returned the top result as the assigned classification for that image.

### 4.3.2 Hybrid Search

The hybrid search method combines both keyword and vector search approaches. This technique aims to integrate the strengths of each search type: the precise lexical matching of the keyword search and the broader semantic understanding provided by vector search. By blending both methods, the hybrid search seeks to increase the likelihood of retrieving the correct Iconclass classification, especially in cases where either approach alone might be insufficient.

### 4.3.3 Retrieval-Augmented Generation (RAG)

The final advanced method employed was Retrieval-Augmented Generation. For this, the descriptions were used as queries to search the database which returned the top five results. These results, combined with the original description, were passed to the LLM, which was prompted to select the best matching Iconclass entry from the top five results. This approach allowed the model to refine its selection by considering multiple closely related results, helping to address cases where the top result might not be the most accurate match. By incorporating retrieval and generation, this method added an additional layer of analysis to improve classification outcomes.

---

40. The OpenAI embedding model was used via Weaviate's text2vec_openai OpenAI integration.

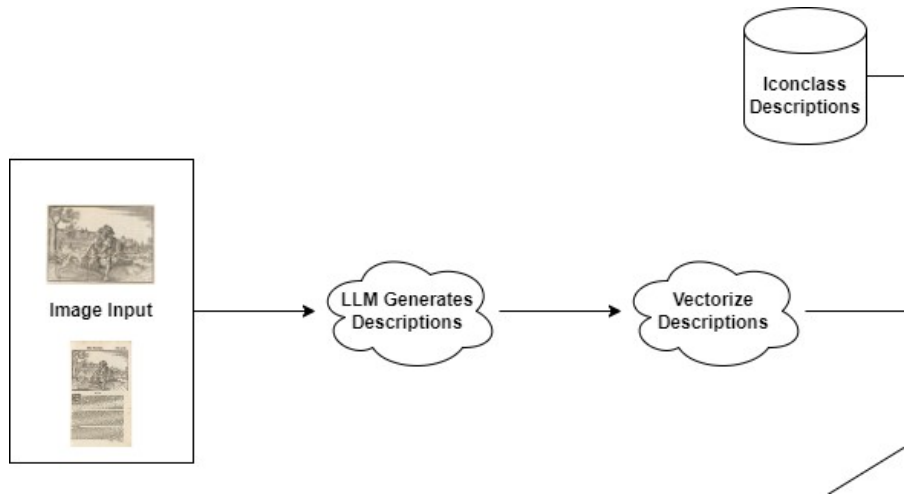

***Fig. 3:*** *Diagram of Methodology*

## 5. Model Evaluation

In this section, I outline the methodology and scoring system used to evaluate model performance. Given the hierarchical and nuanced nature of Iconclass, this evaluation focuses not only on perfect matches but also on partial matches, which play a crucial role in identifying broader themes in the images. To assess performance, I used a combination of traditional precision, recall, and F1 score metrics adapted for hierarchical classification, as well as a more nuanced weighted scoring system that captures different levels of classification accuracy.

### 5.1 Hierarchical Precision, Recall, and F1 Score

Since Iconclass codes are hierarchical, a prediction can be partially correct by matching some of the higher levels of the classification without perfectly matching the lower levels. This partial correctness is important for the project, as correctly identifying the parent levels can still provide valuable insight into the broader theme of the image. For example, knowing an image depicts "the story of David and Goliath (1 Samuel 17)" (71H14) is useful even if a more accurate classification would be "David slings a stone at Goliath's forehead" (71H1442), which is two levels deeper.

To evaluate model performance at each level of the classification, I wrote a script using the Iconclass Python package[41] to compare the predicted classification to the ground truth by looking up each code's parent levels, calculating the total number of levels for each, and then determining how many levels match between them.

Precision in this context measures how many of the predicted levels are correct out of the total number of levels in the prediction:

$$\text{Precision} = \frac{\text{Number of Correct Prediction Levels}}{\text{Total Prediction Levels}}$$

Recall measures how many levels of the ground truth are correctly predicted:

$$\text{Recall} = \frac{\text{Number of Correct Prediction Levels}}{\text{Total Ground Truth Levels}}$$

F1 Score balances precision and recall by calculating the harmonic mean to give an overall measure of the model's accuracy, calculated as:

$$\text{F1-Score} = 2 \times \frac{\text{Precision} \times \text{Recall}}{\text{Precision} + \text{Recall}}$$

Each metric is calculated for each image to be used in an average score for the whole model. This approach provides a quantitative evaluation of how well the model captures not only the specific details but also the broader themes represented by higher-level Iconclass codes.

**5.2 Weighted Scoring System**

To account for varying degrees of classification accuracy, a weighted scoring system was applied that reflects the hierarchical nature of Iconclass codes and offers more nuance than the traditional metrics. This system categorizes predictions into five distinct match types, each assigned a base score according to its relative accuracy:

- **Full Match (100 points)**: The predicted code exactly matches the ground truth, capturing both broad and specific details. A full match represents the ideal classification scenario.

- **Extra Match (90 points)**: The prediction includes all of the classification levels of the ground truth but with additional levels. The minor deduction reflects the possibility that extra levels may not be accurate.

- **Partial Match Type A (85 points)**: The prediction includes fewer levels than the ground truth, but all included levels are correct. This type of match still captures the primary iconographic theme, which is crucial for comparative analysis. The score reflects the high utility of this match, despite the lack of complete detail.

- **Partial Match Type B (70 points)**: The prediction includes fewer levels than the ground truth, and only some of the predicted levels are correct. While useful, this match misses important aspects of the image's classification, justifying a lower score.

- **Partial Match Type C (60 points)**: The prediction includes equal or more levels than the ground truth, but not all levels are correct, introducing incorrect or misleading details, which could distort the classification.

- **No Match (0 points)**: None of the predicted levels match the ground truth.

To further refine the scoring system, there are percentage-based adjustments for partial matches. For each partial match, the base score is multiplied by the percentage of levels that match the ground truth. For example, if the ground truth has five levels and the prediction correctly matches three of them, the base score is adjusted to reflect a 60% match. This percentage-based adjustment enables a more detailed picture of the model's ability to approximate the ground truth, even when a perfect match is not achieved. When a prediction includes incorrect levels, a deduction is applied to the score, with a cap set to prevent the score from falling below 50% of the base score. This ensures that partial matches still receive credit for capturing key elements of the classification, even if errors are present.

---

41. https://pypi.org/project/iconclass/

The weighted scoring system, combined with hierarchical precision, recall, and F1 scores, offers a flexible and nuanced framework for evaluating model performance. This approach captures both the accuracy and utility of the model's predictions even if they miss some finer details.

## 6. Results

| Query | Image | Database | Average of Weight Scores | Avg Precision | Avg Recall | Avg F1 Score |
|---|---|---|---|---|---|---|
| image | illustration | image | 31.26168 | 0.3763 | 0.2844 | 0.3017 |
| keyword | illustration | basic | 38.36353 | 0.435 | 0.4909 | 0.4561 |
| keyword | illustration | hierarchical | 42.10763 | 0.444 | 0.5188 | 0.4728 |
| keyword | page | basic | 49.3544 | 0.5995 | 0.6427 | 0.6137 |
| keyword | page | hierarchical | 52.04223 | 0.6001 | 0.6705 | 0.6272 |
| vector | illustration | basic | 55.73697 | 0.5806 | 0.6613 | 0.6131 |
| vector | illustration | hierarchical | 58.29538 | 0.522 | 0.6661 | 0.5801 |
| vector | page | basic | 66.91106 | 0.7219 | 0.8054 | 0.7563 |
| vector | page | hierarchical | 64.70515 | 0.6057 | 0.7524 | 0.6654 |
| hybrid | illustration | basic | 53.90327 | 0.5764 | 0.6606 | 0.6095 |
| hybrid | illustration | hierarchical | 61.3907 | 0.568 | 0.7229 | 0.6299 |
| hybrid | page | basic | 63.62535 | 0.7084 | 0.7891 | 0.7404 |
| hybrid | page | hierarchical | 69.59164 | 0.6801 | 0.8351 | 0.7429 |
| RAG-vector | page | basic | 77.39061 | 0.7895 | 0.8808 | 0.8248 |
| RAG-hybrid | page | hierarchical | 72.33898 | 0.7074 | 0.8401 | 0.7609 |

***Fig. 4:*** *Model evaluation metrics*

| Query | Image | Database | Extra Match | Full Match | Partial Match - Insufficient Depth with Complete Matching | Partial Match - Insufficient Depth with Partial Matching | Partial Match - Sufficient or Excessive Depth with Partial Matching | No Match |
|---|---|---|---|---|---|---|---|---|
| image | illustration | image | 31 | 53 | 112 | 62 | 111 | 221 |
| keyword | illustration | basic | 95 | 36 | 9 | 35 | 252 | 164 |
| keyword | illustration | hierarchical | 114 | 40 | 8 | 32 | 267 | 130 |
| keyword | page | basic | 102 | 75 | 33 | 30 | 211 | 118 |
| keyword | page | hierarchical | 123 | 73 | 18 | 25 | 246 | 84 |
| vector | illustration | basic | 137 | 103 | 1 | 46 | 267 | 37 |
| vector | illustration | hierarchical | 283 | 19 | 1 | 21 | 203 | 64 |
| vector | page | basic | 159 | 148 | 9 | 24 | 208 | 21 |
| vector | page | hierarchical | 297 | 40 | 3 | 13 | 153 | 63 |
| hybrid | illustration | basic | 148 | 71 | 5 | 41 | 285 | 41 |
| hybrid | illustration | hierarchical | 277 | 33 | 2 | 24 | 218 | 37 |
| hybrid | page | basic | 156 | 120 | 13 | 28 | 217 | 35 |
| hybrid | page | hierarchical | 294 | 66 | 7 | 11 | 151 | 40 |
| rag-vector | page | basic | 190 | 199 | 20 | 23 | 123 | 14 |
| rag-hybrid | page | hierarchical | 282 | 97 | 9 | 20 | 125 | 36 |

**Fig. 5:** *Results by Match Type*

### 6.1 Image-Based Search

The image-based search method used the Iconclass AI Test Set for classifying the test images. Across the entire corpus, the model achieved 53 full matches, where the predicted Iconclass code exactly aligned with the ground truth. However, the model struggled significantly with 221 no matches, underscoring the difficulty in identifying correct classifications for many images. The model scored higher among the thematic subset, as there were only three full matches among the biblical illustrations. The weighted score for the 1534 Bible was only 19.5. The average weighted score for the whole corpus was 31.26. The model's precision was 0.4261, with recall and F1-scores of 0.3097 and 0.3017 respectively, illustrating a consistent challenge across the dataset. Overall, the low performance indicates that the image-based search method struggles with accurately classifying early modern religious woodcuts.

## 6.2 Keyword-Based Search

The keyword-based search queried both types of image descriptions—those based on the illustration alone and those based on the full-page context—across both databases. Overall, the search using full-page descriptions with the hierarchical database yielded better results, with 123 extra matches, 73 full matches, and 84 no matches. The keyword-based search outperformed the image-based search across all test sets. The precision, recall, and F1-scores were consistently higher for keyword queries, highlighting the advantage of leveraging textual descriptions to capture the contextual and thematic nuances of the illustrations. The highest scoring keyword search had an F1-score of 0.6272, compared to 0.3017 for the image search. These results establish a valuable baseline for comparing more advanced search methods.

## 6.3 Results: Vector Search and Hybrid Search

### 6.3.1 Total Corpus Results

Across the corpus, full-page, contextual descriptions consistently outperformed those derived from the illustrations alone, particularly in weighted scores and F1 scores, regardless of the search method used. For instance, with the hierarchical database and hybrid search, full-page descriptions achieved an average weighted score of 69.59 and an F1 score of 0.7429, significantly higher than the illustrations alone, which recorded an average weighted score of 61.39 and an F1 score of 0.6299.

The hierarchical database yielded better results in general. In particular, full pages classified with the hierarchical database and hybrid search achieved an average recall of 0.8351 compared to 0.7891 for the basic database. However, the basic database proved better with vector search. For example, with full pages and vector search, the basic database led to a slightly higher F1 score (0.7563) compared to hybrid search (0.7404), indicating fewer incorrect or extraneous levels.

Hybrid search outperformed vector search when using the hierarchical database. For full pages with the hierarchical database, hybrid search produced an average F1 score of 0.7429, while vector search lagged behind at 0.6654. This indicates that hybrid search, leveraging both lexical and semantic similarities, provided a more balanced classification approach for complex, multi-level Iconclass codes. Conversely, for the basic database, vector search occasionally performed slightly better in terms of precision, though the differences were marginal.

### 6.3.2 Subset Results

In the evaluation of illustrations from the 1534 and 1551 Luther Bibles, full-page descriptions consistently outperformed those based on the illustrations alone. In the 1534 edition, full pages with hybrid search achieved an F1 score of 0.7696, compared to 0.5234 for cropped images. The basic database also outperformed the hierarchical database, particularly for full pages. Hybrid search slightly outperformed vector search, particularly for full pages, with the 1551 edition achieving an F1 score of 0.7978 using hybrid search compared to 0.7749 with vector search.

In contrast, the illustration descriptions slightly outperformed full-page descriptions in the selected thematic illustrations, achieving higher scores for both vector and hybrid searches, particularly when paired with the hierarchical database. The illustrations alone with vector search reached an F1 score of 0.7826, while full pages performed slightly lower. The hierarchical database consistently showed stronger results compared to the basic database. Additionally, vector search outperformed hybrid search with the illustrations, although hybrid search worked better for full pages.

When comparing the results of vector and hybrid search methods to the image-based and keyword-based search methods, the improvements are substantial. The vector and hybrid search approaches achieved notably higher weighted and F1 scores, particularly when using full-page images and the hierarchical database. For example, the highest F1 score for the total corpus using hybrid search and full pages was 0.7429, significantly surpassing the F1 score of 0.3017 achieved with image-based search. Similarly, the hybrid search method consistently produced more accurate classifications than the keyword-based search.

While the keyword-based search yielded an F1 score of 0.6272 with full-page descriptions, the vector search achieved much stronger performance with scores reaching 0.7563. These results underscore the power of combining semantic and lexical search techniques, marking a significant improvement over the earlier image and keyword-based methods.

**6.4 Results: RAG Implementation**

The implementation of Retrieval-Augmented Generation resulted in notable improvements across the corpus. By allowing the LLM to consider the top five search results, it was able to make more informed decisions. This method was applied to the two top performing models and showed gains in average weighted scores, precision, recall, and F1 scores. The inclusion of multiple classification options helped correct initial ranking errors and reduce misclassifications.

After implementing RAG with the model using hybrid search on the hierarchical database with full pages, the average weighted score increased from 69.59 to 72.34, and the F1 score rose from 0.7429 to 0.7609. The modest gains suggest the model already excelled in retrieving relevant levels, providing fewer opportunities for improvement.

The most significant improvement was observed in the model searching the basic database with the full-page descriptions using vector search. This was the second-best model overall based on the weighted score and best overall F1 score, but when combined with RAG, it outperforms the other model. The average weighted score jumped from 66.91 to 77.39, while the F1 score increased substantially from 0.7563 to 0.8248. These gains reflect a balanced enhancement of both precision and recall, indicating that RAG not only improved the model's ability to find relevant classifications but also reduced false positives, leading to a higher overall performance.

When comparing match types, RAG reduced the number of no matches, indicating that the LLM was more consistently able to assign appropriate classifications. In the Pages, Basic Database, Vector Search model, full matches increased from 148 to 199, signaling a significant improvement in the model's accuracy. Moreover, partial matches that have sufficient or excessive depth decreased significantly from 208 to 123. However, extra matches remained substantial, particularly in the hierarchical database model, reflecting the LLM's tendency to include more detailed but sometimes incorrect levels.

## 7. Discussion

### 7.1 Vector and Hybrid Search Methods: What Worked and Why?

The hybrid search method generally outperformed vector search when paired with the hierarchical database. This result aligns with the expectation that hybrid search benefits from leveraging both lexical matches and semantic similarities. Because the hybrid method considers keyword matching as a part of its score, the hierarchical database provided more opportunities for a match. This was especially evident when working with full-page images, where the surrounding text provided additional context, allowing for more accurate classification.

However, vector search proved more effective when using the basic database, particularly with the illustrations images. The basic database's shorter descriptions seemed to help prevent over-complication. This method worked well for simpler or more iconic images which do not need the surrounding context to capture essential visual features efficiently.

One notable challenge across both search methods was the prevalence of extra matches, especially when using the hierarchical database. This suggests that while the model often captured the correct core classification, it struggled to limit its depth, including additional and sometimes incorrect hierarchical levels. Manual inspection revealed many of the extra levels provided wrong or debatable detail.[42] Partial matches were also frequent, indicating that the model often captured the broader themes of an image but failed to correctly classify the finer details.

**7.2 RAG Method: How It Improved the Results and What It Tells Us**

The implementation of RAG marked a noticeable improvement over the vector and hybrid search methods. By feeding the top five search results into an LLM, the model could make a more informed decision when selecting the best Iconclass. This mitigated the impact of ranking errors from the initial search methods, as the best match was not always the top-ranked result. The LLM's decision-making ability helped reduce the number of no matches and partial matches while increasing full matches, indicating more precise and confident classifications.

The greatest gains from RAG were observed in the Pages, Basic Database, RAG-Vector model, which saw significant improvements in average weighted scores, precision, and recall. The precision improvements suggest that the LLM could filter out incorrect levels and make more accurate classifications. The increase in recall shows that the LLM retained the ability to capture relevant classification levels, even as it refined the depth of its classifications. However, even with RAG, the persistent issue of extra matches remained, suggesting that further refinement is needed to handle over-specification.

**7.3 How many levels are enough?**

One way to mitigate incorrect levels in both extra and partial matches is to examine what classification depth is the most accurate. In both RAG models the average number of classifications per prediction was just over six. If we look at the number of correct matches level by level, there is a small drop-off between levels three and four, and a large drop-off at level five. In the RAG model with vector search on the basic database, level three matches are 94% accurate, with a decrease to 84% at level four, and a large drop-off at level 5 (65%). The RAG model using hybrid search with the hierarchical database is similar though slightly lower. This suggests a maximum prediction depth of four levels could improve classification accuracy.

---

42. Thanks to my research assistant, Lena Böse, for her help examining such matches.

| Query | Image | Database | Level | Objects | Percent |
|---|---|---|---|---|---|
| RAG - Vector | page | basic | 1 | 554 | 97.36 |
| | | | 2 | 552 | 97.01 |
| | | | 3 | 538 | 94.55 |
| | | | 4 | 479 | 84.18 |
| | | | 5 | 368 | 64.67 |
| | | | 6 | 127 | 22.32 |
| | | | 7 | 45 | 7.91 |
| | | | 8 | 15 | 2.64 |
| | | | 9 | 2 | 0.35 |
| RAG - Hybrid | page | hierarchical | 1 | 532 | 93.50 |
| | | | 2 | 524 | 92.09 |
| | | | 3 | 510 | 89.63 |
| | | | 4 | 456 | 80.14 |
| | | | 5 | 343 | 60.28 |
| | | | 6 | 123 | 21.62 |
| | | | 7 | 41 | 7.21 |
| | | | 8 | 11 | 1.93 |

*Fig. 6: Accuracy by classification level*

Another method is to recalculate evaluation metrics after truncating the predictions by one level at a time. In the RAG model with vector search on the basic database, truncating the last level from each predicted Iconclass raises the model's precision from 0.7909 to 0.8694. Truncating two or three levels raises the precision further to 0.9246 and 0.9555, respectively. This still leaves an average depth of five or four levels. However, although the precision increases, the recall drops significantly. The reason for the precision increase with truncation is that as you remove levels from the predicted Iconclass code, you are effectively "generalizing" the predictions. As a result, the truncated predictions are more likely to match the ground truth at broader hierarchical levels. Truncating a code removes specific details, which reduces the chance of being wrong in those details. Thus, the system becomes more conservative and often correct at the broader, higher-level categories.

Recall measures how many of the ground truth codes are captured by the predictions. As

25you truncate levels, the predicted codes become more generalized, meaning fewer of the specific details of the ground truth are matched. Although these broader categories may still be relevant, they fail to capture the more specific information contained in the ground truth, leading to a lower recall score. These results show that truncation allows for a trade-off between precision and recall.

Therefore, determining the appropriate number of levels depends on the specific goals of the project. If broad classifications are sufficient, setting a maximum depth of four levels or applying truncation can effectively balance precision and recall, while still retaining relevant thematic information without over-specification. It may be that recall and thus F1 scores are not the best evaluation metrics for Iconclass, as low-level, broader classifications are still useful.

| Query | Image | Database | Levels Truncated | Precision | Recall | F1 Score | Avg Levels |
|---|---|---|---|---|---|---|---|
| RAG - Vector | pages | basic | 0 | 0.7909 | 0.8823 | 0.8262 | 6.08 |
| | | | 1 | 0.8694 | 0.8126 | 0.8302 | 5.08 |
| | | | 2 | 0.9246 | 0.6972 | 0.7819 | 4.08 |
| | | | 3 | 0.9555 | 0.5462 | 0.6768 | 3.08 |
| | | | 4 | 0.9463 | 0.37 | 0.5089 | 2.08 |
| RAG - Hybrid | pages | hierarchical | 0 | 0.7137 | 0.8476 | 0.7677 | 6.4 |
| | | | 1 | 0.8139 | 0.8168 | 0.8067 | 5.4 |
| | | | 2 | 0.8789 | 0.7214 | 0.7815 | 4.4 |
| | | | 3 | 0.9131 | 0.5821 | 0.6968 | 3.4 |
| | | | 4 | 0.9275 | 0.4163 | 0.5561 | 2.4 |

**Fig. 7**: *Recalculated evaluation metrics by truncation level*

## 8. Limitations of the Study

While the methodology of using LLM-generated descriptions in combination with a vector database has produced encouraging results, several limitations affect the precision and reliability of the assigned Iconclass codes.

One key limitation is the dependence on the accuracy of the generated descriptions. Full-page images consistently yielded higher precision in Iconclass predictions than the illustration images, which lacked contextual information necessary for accurate classification. Cropped images performed well only when they depicted widely recognized scenes, like the crucifixion, where the context is more implicit.

The LLM's proficiency with religious and biblical imagery also presented a limitation. While the model was well-suited to this specific dataset, its strength in describing religious content may not transfer effectively to secular or non-religious illustrations. Future applications involving different genres may see a drop in performance due to this reliance on the LLM's specialized understanding of biblical motifs.

There are also issues with the books that can disrupt model accuracy. Pages might contain more than one image, requiring a unique solution to avoid LLM confusion.[43] Another challenge arose from misalignment between illustrations and their accompanying text. In the 1534 Bible, Isaiah begins with an illustration from chapter six. It was also common for books to have composite or multi-scene illustrations, where multiple moments from a narrative might coexist in a single image, such as Adam and Eve at the tree of life alongside their expulsion from Eden. In such cases, the LLM might prioritize a different part of the image than the ground truth, leading to mismatches and thus lower scores, even though both are correct. Accordingly, this method was limited by its restriction to assigning only one Iconclass code per image. Many early modern illustrations contain multiple layers of symbolism, and reducing these complex images to a single classification could miss important aspects of the narrative.

The reuse of illustrations across different texts also introduced ambiguity. In early modern printing, woodcut images were sometimes repurposed across unrelated stories, potentially confusing the LLM or the resulting Iconclass predictions. Moreover, Iconclass itself does not cover every biblical scene with equal specificity, and the subjective nature of the system means that different annotators could assign different codes to the same image or broader classifications may be chosen when specific matches are unclear. This subjectivity in ground truth coding may have penalized some of the model's predictions, even when both the prediction and the human-assigned code were reasonable interpretations.

Overall, while this method demonstrates significant promise for automating the classification of early modern religious illustrations, challenges related to complexity and context remain to be addressed.

## 9. Conclusion

One of the main takeaways from this project is the importance of adapting techniques to the nature of the dataset. The combination of RAG, vector search, and the basic Iconclass database works exceptionally well. The success of the full-page descriptions has demonstrated the importance of context in generating accurate image descriptions for classification. We could supplement the lack of descriptive metadata with the information gleaned from the full pages. Moreover, it proved that the LLM can successfully read early modern German text printed in blackletter type. Similarly, restricting the Iconclass database to relevant categories limited the scope of possible errors.

One immediate avenue for improvement would be to fine-tune an LLM specifically for this task. Fine-tuning would likely improve the model's precision in generating more tailored image descriptions. Also, experimenting with alternative LLMs, descriptive ontologies, or refined prompts to generate more specific image descriptions could enhance results.

Another option would be to experiment with different embedding models and distance metrics. The performance of vector searches can vary significantly with different embeddings. Future work could experiment with different embedding models to see if they yield better results.

When using hybrid search, adjusting the weights between the keyword and vector search methods may yield better results. For example, placing more emphasis on vector similarities for well-defined images or emphasizing keywords in cases where textual cues are stronger might enhance performance.

---

43. Such pages were excluded from the test set.

Another area for improvement would be optimizing RAG retrieval strategies. Currently, only the top five results were retrieved in RAG, but expanding this to the top ten or experimenting with combinations of results could improve classification accuracy. In one example, the correct Iconclass was the sixth result, which was not passed to the LLM. Also, by including parent levels of Iconclass codes during RAG retrieval, it might reduce the problem of over-specification. Lastly, since RAG processing is computationally expensive, it was applied only to the top two performing models. Applying RAG to all models could reveal if another model might outperform the others when RAG is included. This would allow a more comprehensive comparison and ensure the most effective model is being used.

The next phase of my research will involve applying the top-performing models across the entire corpus of 120,000 images, expanding the scope of the classification effort. I will implement depth-level truncation, limiting the classification to four or five levels. Given the structure of biblical stories, particularly in the Old Testament, three classification levels are often sufficient to capture the main narrative, book, or scene depicted. This ensures that the classifications remain broad enough to capture the major biblical stories while avoiding unnecessary over-specification. However, to allow for future manual inspections or refinements, I will also save the original outputs before truncation.

This project demonstrates that the method of applying RAG with vector search across a basic Iconclass database is a highly effective tool for automatic Iconclass classification. Using descriptions generated from full-page images, this approach achieved 92% precision at four classification levels and 87% at five levels—a significant improvement in prediction accuracy over traditional keyword search methods (52%). The integration of full-page context into the image description process greatly enhanced the quality of the classification, confirming that text-based methods can succeed even in image-driven tasks.

What makes this methodology particularly advantageous is its multi-modal nature, which combines images and text as inputs. This process did not require any fine-tuning or the creation of specialized training sets for images, making it easy to implement and practical for large-scale projects. While fine-tuning may further improve results, the high accuracy levels achieved without it demonstrate the robustness of this approach.

This study also highlights the growing relevance of multi-modal large language models in domains traditionally dominated by convolutional neural networks and object detection methods. The success of LLMs in tasks like this, previously seen as beyond their scope, underscores their potential in iconographic studies and beyond. Furthermore, the use of vector databases to manage and query hierarchical Iconclass data showcases the scalability and flexibility of these tools. By integrating vector searches with RAG, the model benefits from both structured data retrieval and flexibility in ranking results, allowing for nuanced classification.

Additionally, the results demonstrate that Iconclass is not a binary task—a partial match at higher levels can still provide useful insights, even if deeper levels of classification are not fully accurate. By limiting the depth of classification, when necessary, this method successfully handles broader tasks while mitigating over-specification.

In summary, this method offers a powerful, adaptable, and resource-efficient solution for large-scale image classification, highlighting the potential of LLMs and vector databases for future digital humanities projects.